\documentclass[apl, reprint, showpacs,citeautoscript, superscriptaddress]{revtex4-1}
\usepackage{dcolumn}
\usepackage{bm}
\usepackage[T1]{fontenc}
\usepackage{amsmath,amsfonts,amssymb}
\usepackage{graphicx}
\usepackage{amssymb}
\usepackage{epstopdf}
\usepackage{color}

\begin{document}

\title{Towards bright and pure single photon emitters at 300 K based on GaN quantum dots on silicon}

\author{Sebastian Tamariz}
\email{sebastian.tamariz@epfl.ch}
\affiliation{Institute of Physics, \'Ecole Polytechnique F\'ed\'erale de Lausanne, EPFL, CH-1015 Lausanne, Switzerland}
\author{Gordon Callsen}
\affiliation{Institute of Physics, \'Ecole Polytechnique F\'ed\'erale de Lausanne, EPFL, CH-1015 Lausanne, Switzerland}
\author{Johann Stachurski}
\affiliation{Institute of Physics, \'Ecole Polytechnique F\'ed\'erale de Lausanne, EPFL, CH-1015 Lausanne, Switzerland}
\author{Kanako Shojiki}
\affiliation{Institute of Physics, \'Ecole Polytechnique F\'ed\'erale de Lausanne, EPFL, CH-1015 Lausanne, Switzerland}
\affiliation{Current address: Graduate School of Engineering, Mie University, Tsu 514-8507, Japan}
\author{Rapha\"el Butt\'e}
\affiliation{Institute of Physics, \'Ecole Polytechnique F\'ed\'erale de Lausanne, EPFL, CH-1015 Lausanne, Switzerland}

\author{Nicolas Grandjean}
\affiliation{Institute of Physics, \'Ecole Polytechnique F\'ed\'erale de Lausanne, EPFL, CH-1015 Lausanne, Switzerland}

\begin{abstract}

Quantum dots (QDs) based on III-nitride semiconductors are promising for single photon emission at non-cryogenic temperatures due to their large exciton binding energies. Here, we demonstrate GaN QD single photon emitters operating at 300 K with $g^{(2)}(0) = 0.17 \pm 0.08$ under continuous wave excitation. At this temperature, single photon emission rates up to 6\,$\times$\,10$^6$\,s$^{-1}$ are reached while $g^{(2)}(0) \leq 0.5$ is maintained. Our results are achieved for GaN QDs embedded in a planar AlN layer  grown on silicon, representing a promising pathway for future interlinkage with optical waveguides and cavities. These samples allow exploring the limiting factors to key performance metrics for single photon sources, such as brightness and single photon purity. While high brightness is assured by large exciton binding energies, the single photon purity is mainly affected by the spectral overlap with the biexcitonic emission. Thus, the performance of a GaN QD as a single photon emitter depends on the balance between the emission linewidth and the biexciton binding energy. We identify small GaN QDs with an emission energy in excess of 4.2\,eV as promising candidates for future room temperature applications, since the biexciton binding energy becomes comparable to the average emission linewidth of around 55\,meV.

%





\end{abstract}

\maketitle

Quantum dots (QDs) based on III-V semiconductors have attracted a lot of attention for their use as non-classical light sources, with the single photon source being the simplest and most elemental representative. Such a source of single photons should be as bright as possible, while retaining a high single photon purity \cite{Senellart2017, Schweickert2018}. However,  key metrics for such QD-based single photon sources are usually achieved at cryogenic temperatures with the seminal In(Ga)As/(Al)GaAs system \cite{Dou2008, Dusanowski2015, Cavigli2012}. Identifying a material platform that can enable sufficiently performant single photon sources up to room temperature remains a challenging quest. In this respect, the main contenders are point defects in wide-bandgap semiconductors (2D materials \cite{He2015} and bulk semiconductors \cite{Castelletto2014, Morfa2012}), nitrogen and silicon vacancies in diamond \cite{Aharonovich2011}, as well as semiconductor QDs \cite{Michler2000, Tribu2008, Lin2017}. It would be advantageous to employ a material system with high integrability into a suitable photonic environment that offers epitaxial control. In this regard, III-nitrides offer a unique possibility as bipolar doping can be achieved, foreign and homoepitaxial substrates are available, and growth and processing techniques are well established, leading to their widespread implementation at an industrial scale for solid state lighting. 


III-nitrides have shown promising advances in terms of single photon emission (SPE) by employing GaN/AlN \cite{Santori2005, Kako2006}, GaN/AlGaN \cite{Holmes2014, Arita2017} and InGaN/GaN QDs \cite{Kremling2012, Jemsson2015, Wang2017, Gacevic2017, Cho2018}. Furthermore, SPE at temperatures as high as 350 K \cite{Holmes2016} and two-photon emission up to 50 K \cite{Callsen2013} have been demonstrated. The progress towards room temperature operation is directly linked to the exciton-phonon coupling. With rising temperature the phonon bath becomes increasingly populated. Optical and acoustic phonon populations can lead to effects such as charge carrier escape from the trapping potential of the QD reducing the brightness \cite{Holmes2016, Dou2008, Dusanowski2015, Cavigli2012} and emission linewidth broadening \cite{Krummheuer2002, Rol2007, Holmes2013, Ostapenko2012}. The latter, is directly linked to the process of dephasing \cite{Callsen2013}, which also limits the coherent manipulation of excitonic qubits \cite{Holmes2013, Thoma2016}. For these reasons, the exciton-phonon interaction is generally considered as detrimental for QD-based single photon sources. Interestingly, for the specific case of GaN/AlN QDs  a benefit can be drawn from the particularly strong exciton-phonon coupling \cite{Callsen2015}. H\"onig \emph{et al}. \cite{Honig2014} have shown that the biexciton cascade in GaN/AlN QDs often requires a phonon-mediated spin flip process in order to decay into the QD ground state.


In this work we explore key performance metrics of GaN/AlN QDs directly grown on  Si(111) substrate. This choice of substrate would allow for future integration of the QDs into photonic nanostructures, owing to the high etching selectivity between the two material systems \cite{Vico2014, Roland2014, Rousseau2017}. Furthermore, the balance between detrimental and beneficial effects induced by the exciton-phonon coupling is assessed. Thus, GaN/AlN single QDs are shown to be bright single photon sources with count rates at the sample surface on the order of 10$^6$ - 10$^7$ s$^{-1}$ at room temperature. Power-dependent second order autocorrelation function measurements ($g^{(2)}(\tau)$) were carried out at room temperature and the traces examined in the framework of a multiple-level system including the biexciton (XX), higher order excitonic complexes, and spectral diffusion. All probed QDs displayed single photon characteristics at 300 K with an average $g^{(2)}(0)$ = 0.23 and a minimum of $g^{(2)}(0)$ = 0.17 $\pm$ 0.08.







\begin{figure*}
	\centering
	\includegraphics[width= 0.8\textwidth]{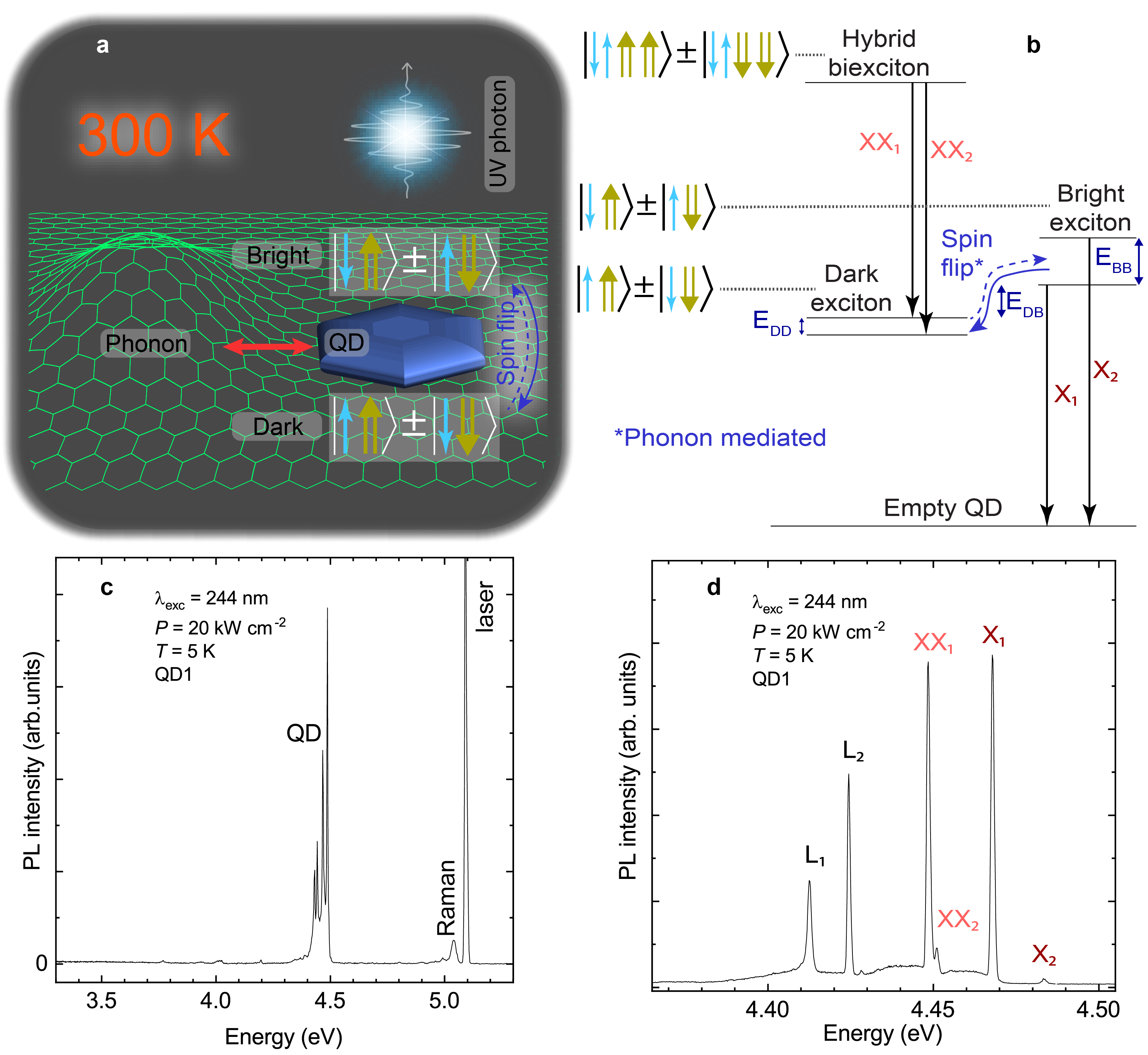}
	\caption{\textbf{a} Schematic illustration of the phonon-assisted spin flip from \emph{dark} to \emph{bright} states. \textbf{b} Energy levels and radiative transitions of a GaN QD, along with the electron ($\uparrow$ /$\downarrow$) and hole ($\Uparrow$ /$\Downarrow$) spins. \textbf{c} Overview $\mu$-PL spectrum recorded at high excitation power of QD1 at low temperature showcasing the background free PL signal. \textbf{d} High-resolution $\mu$-PL spectrum of the same QD recorded at  high excitation power and low temperature.  }
	\label{fig:schema}
\end{figure*}

\section*{Results}

Phonons play an important role in the SPE process in GaN QDs as they provide the energy conservation for the spin flip from \emph{dark} to \emph{bright} excitons as illustrated in Fig. \ref{fig:schema} \textbf{a}. \emph{Dark} excitons are constantly generated in GaN QDs due to the presence of a so-called hybrid biexciton \cite{Honig2014}. In the hybrid biexciton, the two electrons have opposite spins, whereas the holes have parallel spins; which is the energetically stable configuration due to the occurrence of huge hole masses in GaN \cite{Rodina2001}. For this spin configuration, the total spin of the electrons is $S$ = +1/2 -1/2 and that of the holes is $J$ = +(-) 3/2 +(-) 3/2 and hence the total excitonic angular momentum ($M = S + J$) of the hybrid biexciton amounts to $M$ = $\pm$ 3. A diagram of the relevant energy levels, their spin configurations and radiative transitions are displayed in Fig. \ref{fig:schema} \textbf{b}.  The hybrid biexciton-exciton cascade shown in Fig. \ref{fig:schema} \textbf{b} deviates from the classical biexciton cascade; upon decay of the hybrid biexciton in a one-photon process ($\Delta M$ = 1), conservation of angular momentum implies the population of the \emph{dark}-exciton states with $M$ = $\pm$ 2. The \emph{dark} excitons can then transition into \emph{bright} exciton state with $M$ = $\pm$ 1 via a phonon-assisted spin flip process \cite{Labeau2003, Roszak2007, Honig2014}. This process is illustrated in Fig. \ref{fig:schema} \textbf{a} and \ref{fig:schema} \textbf{b}  by the blue dashed arrow.  The intricate physical mechanism behind the phonon-assisted spin flip process is still debated in literature \cite{Woods2002, Johansen2010, Khosla2018}. This latter process depends on the thermal population of phonons and the \emph{dark} to \emph{bright} energy splitting ($E_{\mathrm{DB}}$), see Fig. \ref{fig:schema} \textbf{b}. The spin flip from \emph{bright} to {dark} (solid blue arrow) requires the emission of a phonon and hence does not depend on temperature. As a consequence, the \emph{bright} exciton states are thermally activated by the phonon-mediated spin flip. Further on, due to QD asymmetry, the \emph{bright} states display fine structure splitting (FSS) \cite{Kindel2010} denoted by $E_{\mathrm{BB}}$ in Fig. \ref{fig:schema} \textbf{b}.  A similar, but smaller splitting occurs for the excitonic dark states, which is denoted by $E_{\mathrm{DD}}$ in Fig. \ref{fig:schema} \textbf{b} \cite{Honig2014}.


\begin{figure*}
	\centering
	\includegraphics[width = 0.8\textwidth]{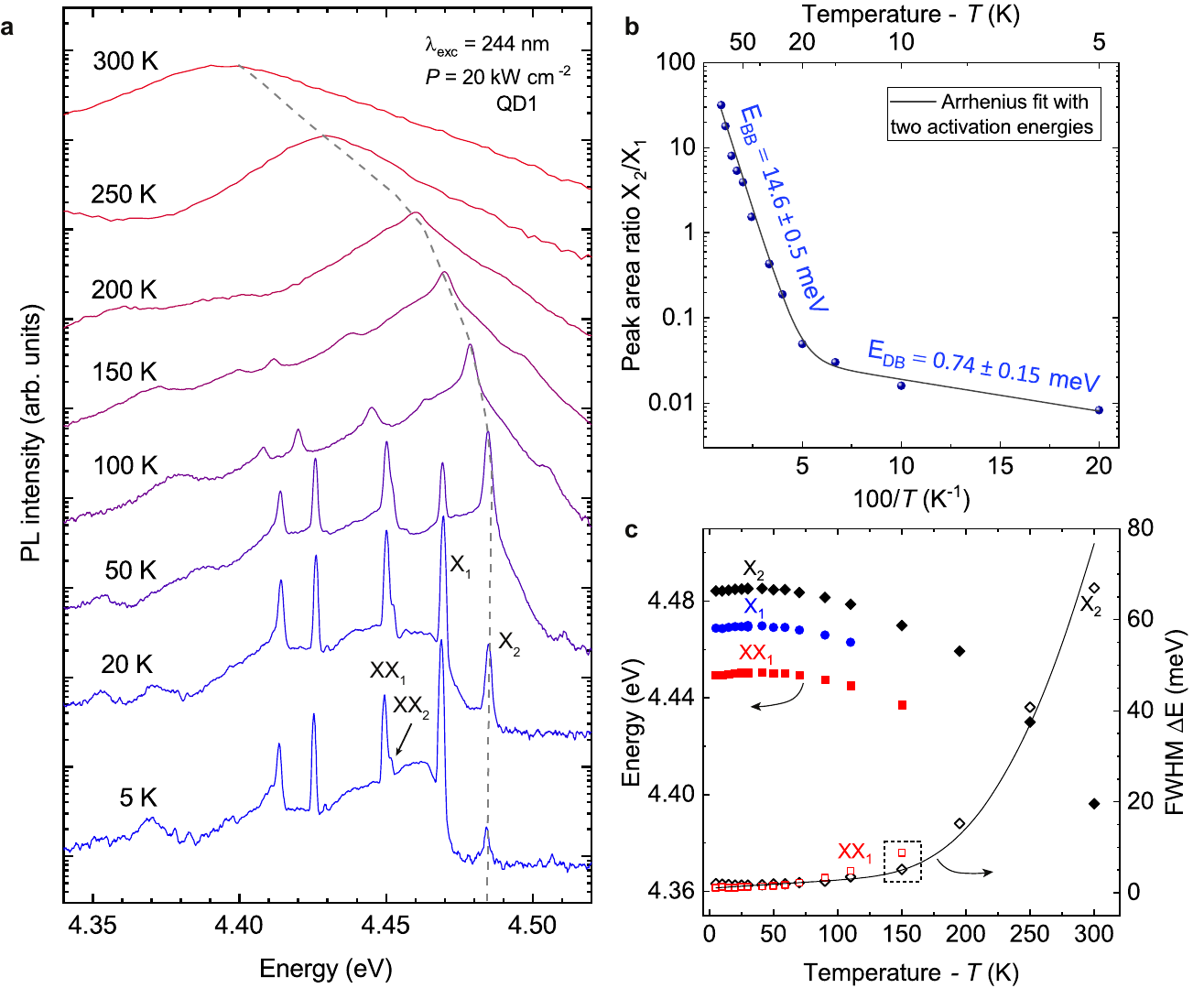}
	\caption{\textbf{a} $\mu$-PL spectra of a single GaN QD as a function of temperature (vertically shifted for clarity). The dashed gray line is a guide to the eye. \textbf{b} Arrhenius plot of the peak area ratio of the bright excitons fitted with two activation energies that approximate the \emph{bright} and \emph{dark} FSS. \textbf{c} Evolution of the emission energy and linewidth of the optical transitions as a function of temperature. The solid line is a fit based on equation \ref{eq:linewidth}.}
	\label{fig:temp}
\end{figure*}

A recurring feature of the optical signature of single GaN QDs at low temperature is the presence of several peaks, of which four peaks dominate in intensity (see further examples in the supplementary material). A broad energy range, low temperature micro-photoluminescence ($\mu$-PL) spectrum of a typical GaN QD (QD1) is displayed in Fig. \ref{fig:schema} \textbf{c} showcasing the absence of background emission, along with a high-resolution spectrum in Fig. \ref{fig:schema} \textbf{d}. Some of these transitions were identified within the framework of the hybrid biexciton. The radiative decays of the hybrid biexciton into the exciton's \emph{dark} states (XX$_1$ and XX$_2$) and the exciton's \emph{bright} states (X$_1$ and X$_2$) are labeled in the $\mu$-PL spectrum shown in Fig. \ref{fig:schema} \textbf{d} in agreement with the solid, black arrows of Fig. \ref{fig:schema} \textbf{b}. The identification of the lines was confirmed by power- and polarization-dependent measurements (see supplementary material). All lines are linearly polarized, with L$_1$, L$_2$, XX$_1$ and X$_2$ being cross polarized to XX$_2$ and X$_1$. The nature of the two lower energy peaks ($L_1$ and $L_2$) is currently being investigated.  The strong exciton-phonon interaction is already evident in the $\mu$-PL spectrum of Fig. \ref{fig:schema} \textbf{d}, as each peak is accompanied by a broad low energy side-band that is associated to acoustic phonons \cite{Ostapenko2012}. Indeed, the emission characteristics and the radiative decay of the hybrid XX are heavily impacted by the exciton-phonon interaction as illustrated in Fig. \ref{fig:schema} \textbf{a}.




Since the spin flip from the \emph{dark} to \emph{bright} state is phonon mediated \cite{Honig2014, Sallen2009, Fernee2009}, temperature has an  important role in the  intensity of each transition. The $\mu$-PL spectrum of QD1 is plotted in Fig. \ref{fig:temp} \textbf{a} as a function of sample temperature. The excitation power density was set to 20 kW cm$^{-2}$. As the temperature is increased, X$_2$ rises in intensity  and matches X$_1$ at about 40 K. At a temperature of around 150 K the spectrum is completely dominated by X$_2$. The intensity ratio of X$_1$ and X$_2$ was evaluated  with the peak area and plotted as a function of temperature in Fig. \ref{fig:temp} \textbf{b}. The data can be fitted with an Arrhenius equation and two activation energies, these energies approximate the \emph{dark} to \emph{bright} energy difference ($E_{DB}$) and the \emph{bright} ($E_{BB}$) FSS \cite{Honig2014} introduced in Fig.\,\ref{fig:schema}.

\begin{figure}
	\centering
	\includegraphics[width = 0.49\textwidth]{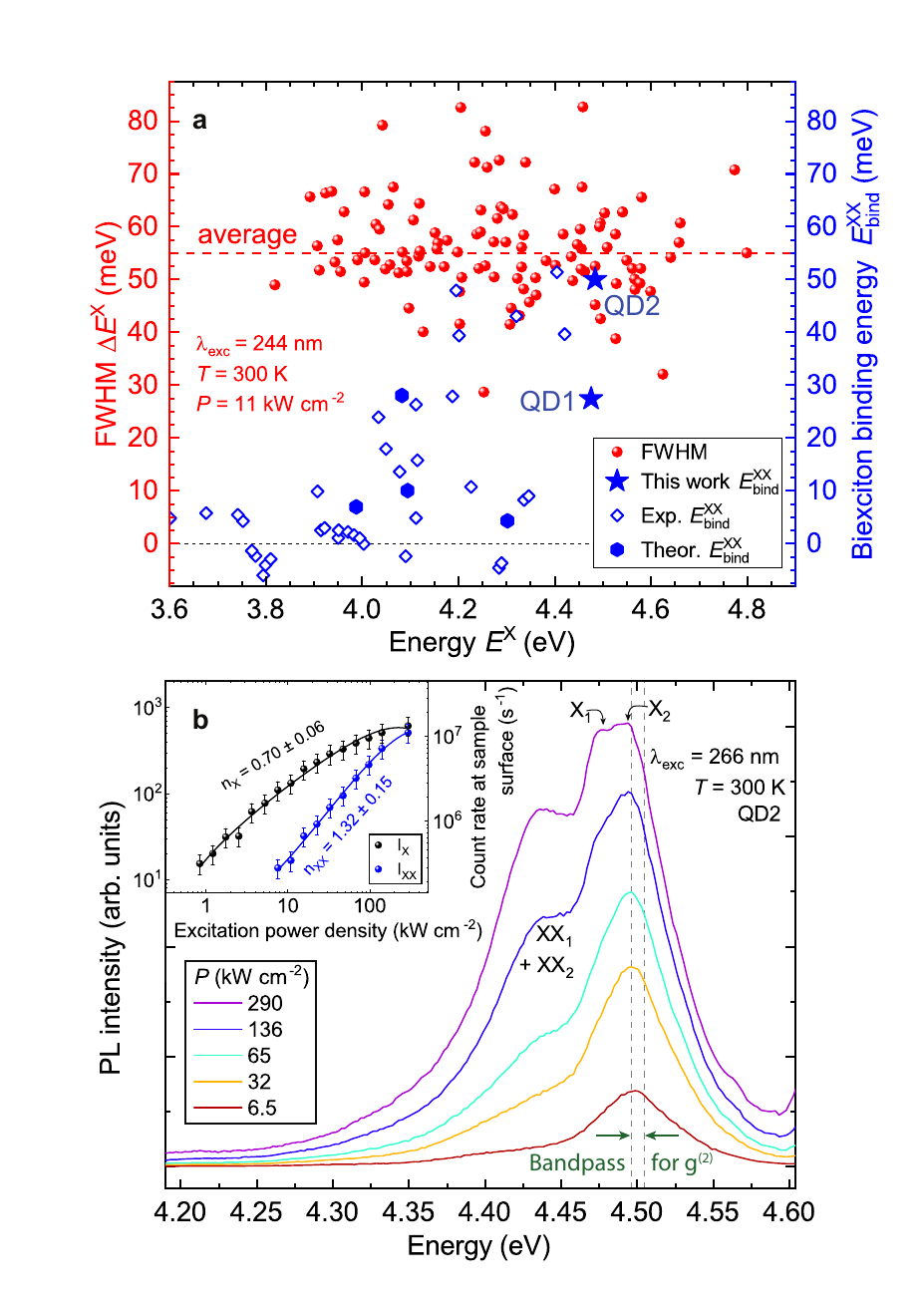}
	\caption{\textbf{a} FWHM of the PL of single GaN QDs at 300 K (red symbols) and biexciton binding energies found in the literature \cite{Holmes2019, Honig2014, Simeonov2008} and of the QDs presented in this work (blue symbols). \textbf{b} Excitation-power-dependent $\mu$-PL spectra of QD2 exhibiting X$_{(1/2)}$ and XX$_{(1/2)}$ transitions. The bandpass used for the $g^{(2)} (\tau)$-measurements is also marked. Inset: PL integrated intensity and count rate at the sample surface as a function of excitation power density for the main PL transitions. The solid lines are fits to the Poisson distribution (see text). The maximum count rate can reach up to $\approx$ 10$^7$ s$^{-1}$.}
	\label{fig:stat}
\end{figure}

The energy and full width at half maximum (FWHM) of the \emph{bright} states and the biexciton were followed throughout the investigated temperature range and are plotted in Fig. \ref{fig:temp} \textbf{c}. The emission of X$_2$ could be tracked over the whole temperature range. It redshifts by about 90 meV from 5 K to 300 K, which corresponds to that of the AlN matrix material \cite{Guo2001}. Notice an additional blueshift in the temperature range from 5 to 40 K on the order of 1 meV, which could originate from the negative thermal expansion coefficient of AlN at low temperatures \cite{Figge09}. Simultaneously, the spectral line broadens to about 60\,meV. The FWHM ($\Gamma(T)$) was modeled (solid line in Fig. \ref{fig:temp} \textbf{c}) using the following equation \cite{Fischer1997}:
\begin{equation}
\Gamma(T)=\Gamma_{0}+\gamma_A T+\frac{\Gamma_{\mathrm{LO}}}{\exp \left(E_{\mathrm{LO}} / k_{B} T\right)-1},
\label{eq:linewidth}
\end{equation}
where $\Gamma_0$ is the inhomogeneous linewidth at 0 K, $\gamma_A$ is the coupling to acoustic phonons, $\Gamma_{\mathrm{LO}}$ is the coupling to longitudinal optical phonons (LO) and $E_{\mathrm{LO}}$ is the energy of LO phonons. $E_{\mathrm{LO}}$ is readily available from the PL data and amounts to $\approx$ 102 meV. Callsen \emph{et al.} found that the $E_{\mathrm{LO}}$ of GaN QDs scales from the value of unstrained GaN for large QDs towards that of AlN for small QDs. This is due to the exciton-LO-phonon interaction averaging over both the GaN QD and the AlN matrix surroundings \cite{Callsen2015}. Then, the constants $\Gamma_{0}$, $\gamma_A$, and $\Gamma_{\mathrm{LO}}$ were fitted to the data with resulting values of 1 $\pm$ 0.1 meV, 18 $\pm$ 2 $\mu$eV/K, and 3.57 $\pm$ 0.26 eV, respectively. The extracted value $\gamma_A$ is larger than the one obtained by Demangeot \emph{et al.} \cite{Demangeot2009}, suggesting that the coupling strength to phonons is specific to each QD as illustrated by linewidth statistics. In Fig. \ref{fig:stat} \textbf{a}, FWHM statistic at a sample temperature of 300 K is presented. The scatter yields an average value of 55 meV with a standard deviation of 7 meV (based on 112 emission lines). Fluctuating charges surrounding GaN QDs can lead to large inhomogeneous broadenings at low temperatures \cite{Kindel2014}. The impact of fluctuating charges on the FWHM of single GaN QDs is exacerbated  due to the quantum confined Stark effect present in polar III-nitride nanostructures. Hence, the linewidth measured at low temperature depends strongly on the excitonic dipole moment and therefore on the QD emission energy.  As opposed to the low temperature case \cite{Kindel2014}, no clear correlation could be observed between the emission energy and the linewidth, signaling a phonon dominated broadening. For a QD to provide SPE at 300 K, a biexciton binding energy ($E_{bind}^{XX}$) comparable to or larger than  the linewidth is required in order to avoid spectral overlap.  In Fig. \ref{fig:stat} \textbf{a}, $E_{bind}^{XX}$ values found in the literature are plotted along with the two QDs presented in this work \cite{Holmes2019, Honig2014, Simeonov2008}. A trend of increasing $E_{bind}^{XX}$ as a function of emission energy can be deduced. At emission energies above $\approx$ 4.2 eV, $E_{bind}^{XX}$ can become comparable to the linewidth rendering small QDs emitting at higher energies even more appealing.

Excitation-power-dependent $\mu$-PL measurements of QD2 are displayed in Fig. \ref{fig:stat} \textbf{b} at a sample temperature of 300 K along with the bandpass employed for $g^{(2)} (\tau)$ measurements. As the power is increased, an additional peak appears on the low energy side, which we suggest to identify as the XX. The integrated intensity of the peaks ($I_X$ and $I_{XX}$ for the exciton and biexciton, respectively) is marked in the inset of Fig. \ref{fig:stat} \textbf{b}. The scaling of $I_X$ and $I_{XX}$ with excitation power density was modeled with a Poisson distribution $I \propto P^{n_i} e^{-P/P_0}$ \cite{Grundmann1997}, where $P$ is the excitation power density and $n_i$ is the number of excitons present in the QD. A fit to the data yielded $n_X$ = 0.70 $\pm$ 0.06 and $n_{XX}$ = 1.32 $\pm$ 0.15. The value of the exponents ($n_i$) are consistent with previous reports on hybrid biexcitons in  GaN QDs \cite{Honig2014}.

A key feature of a practical single photon source is its brightness, which can be compared to other systems by considering the count rate at the sample surface \cite{Aharonovich2016}. The brightness of the QD was estimated for the selected spectral band pass of 8 meV used during the $g^{(2)} (\tau)$-measurements. The brightness of QD2 was quantified by measuring the detection losses in the optical setup. At a temperature of 300 K and  an excitation power of 6.5 kW cm$^{-2}$, the count rate at the sample surface was estimated to be 1.9 ($\pm$ 0.7) $\times$ 10$^6$\,s$^{-1}$.

\begin{figure*}
	\centering
	\includegraphics[width = 0.8\textwidth]{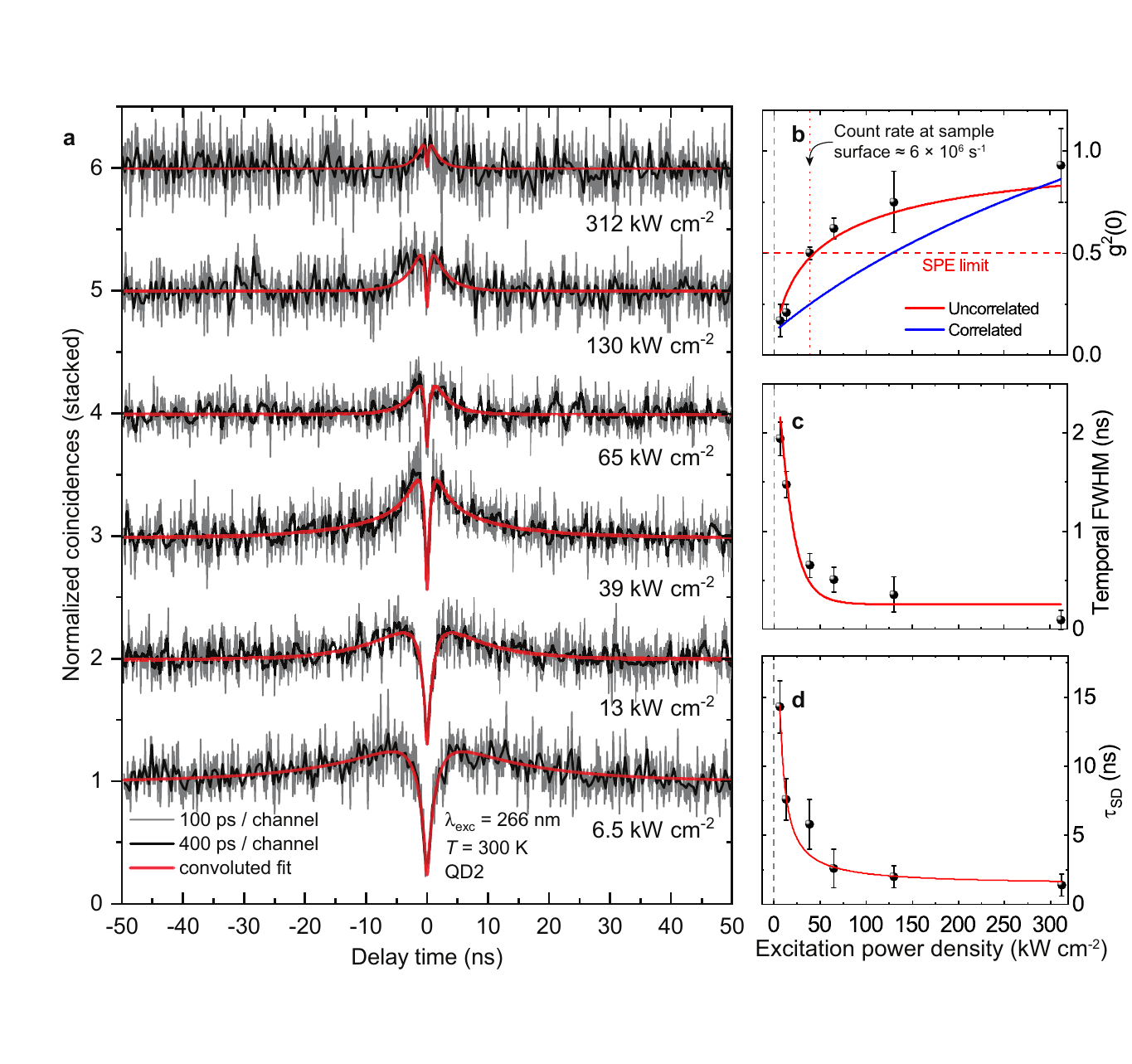}
	\caption{\textbf{a} $g^{(2)} (\tau)$-traces of QD2 at 300 K as a function of excitation power density with channel resolutions of 100 (gray) and 400 (black) ps/channel and convoluted fit (red). \textbf{b} $g^{(2)}(0)$ values as a function of excitation power density. The values surpass the SPE limit for count rates at the sample surface of $\approx$ 6 $\times$ 10$^6$ s$^{-1}$. The loss in single photon purity with rising excitation power density is induced by the XX transition. The solid lines model the corresponding impact of uncorrelated (red) or correlated (blue) photons that perturb the single photon stream of X. \textbf{c} Temporal FWHM of the antibunching peak and fit based on equation \ref{eq:FWHM} (solid red line). \textbf{d} Characteristic spectral diffusion time as a function of excitation power density along with a fit based on $1/\tau_{SD} \propto \sqrt{P}$.}
	\label{fig:g2}
\end{figure*}





The $\mu$-PL emission of these QDs, which is close to background free (see Fig. \ref{fig:schema} \textbf{c}), together with the large $E_{bind}^{XX}$ values and count rates at the sample surface in the 10$^6$ s$^{-1}$ regime,  make these QDs suitable for SPE at 300 K. The performance of QD2 as a single photon source was probed as a function of excitation power density. The results are displayed in Fig. \ref{fig:g2} \textbf{a}. The experimental $g^{(2)}(\tau)$ traces were fitted with a multiple-level model for the biexciton-exciton cascade \cite{Kindel_thesis} that also considers bunching due to spectral diffusion ($g^{(2)}_{SD}(\tau)$) \cite{Sallen2010} convoluted with the instrument response function (IRF):

\begin{equation}
g^{(2)}_{total} (\tau)  = (g^{(2)}_{X}(\tau) \cdot g^{(2)}_{SD}(\tau)) \otimes IRF(\tau),
\end{equation}
where

\begin{equation}
 g^{(2)}_{X}(\tau)  = \exp(\mu e^{-\mid \tau \mid / \tau_d}) (1- e^{-\mid \tau \mid / \tau_d})
 \label{eq:g2tot}
\end{equation}
is the second order autocorrelation function of the exciton, with $\mu = \Pi \cdot \tau_d$, with $\tau_d$ the decay time and $\Pi$ the pump rate. For vanishing excitation powers, $g^{(2)}_{X}(\tau)$ tends to the two-level expression: $g^{(2)}(\tau) = 1 - e^{-\mid \tau \mid / \tau_d}$ \cite{Fox2006}.  Time-resolved PL experiments performed on the QD ensemble at the same emission energy give access to $\tau_d$, while $\mu$ and the spectral diffusion parameters are ascertained by a fit to the data shown in Fig. \ref{fig:g2} \textbf{a}.


A clear room temperature antibunching can be observed at zero time delay proving SPE. At the lowest reported excitation power density, the single photon purity reaches $g^{(2)}(0)$ = 0.17 $\pm$ 0.08.  However, as the excitation power density is increased, $g^{(2)}(0)$ also increases and surpasses the single photon limit ($g^{(2)} (0)$ = 0.5) for an excitation power density of $\approx$ 39 kW cm$^{-2}$ (see Fig. \ref{fig:g2} \textbf{b}). At this excitation power density, the count rate at the sample surface amounts to $\approx$ 6 $\times$ 10$^6$ s$^{-1}$. 

The single photon purity seems to be mainly curtailed by the XX emission, cf. Fig. \ref{fig:stat} \textbf{b}. A quantification of the spectral overlap of the X and XX emission is given in the supplementary material. Although a $E_{bind}^{XX}$ of 47 meV was measured for this specific QD, the XX broadening of this particular QD amounts to almost 80 meV and thus photons from the XX are also present in the investigated spectral window of 8 meV as depicted in Fig. \ref{fig:stat}\textbf{b}. Hence, for rising excitation powers an increasingly prominent luminescence background occurs in the X detection window due to the presence of XX. Thus, the photon stream is contaminated, which limits the achievable $g^{(2)}(0)$ value. However, for a precise quantification of this limitation it is of utmost importance to determine  whether the background luminescence is (a) correlated or (b) uncorrelated to the X emission.

Generally, in the biexciton-exciton cascade, the decay of XX and X should be temporally correlated, leading to a correlation  between the two photon streams linked to XX and X. For the correlated case (a), the resulting $g^{(2)}(0)$ values can be modeled as described in the supplementary material based on a multiexcitonic model, neglecting the \emph{dark} excitons. Practically, as soon as the XX and X emissions overlap in the detection bandpass, all auto- and cross-correlation contributions to   $g^{(2)}(\tau)$ must be summed. As a result of the corresponding theoretical treatment, we obtain the blue, solid line in Fig. \ref{fig:g2} \textbf{b}.

For case (b), the single photon purity can be estimated based on Sallen \emph{et al.} \cite{Sallen2009}:

\begin{equation}
g^{(2)}(0) = 1 - \rho^2 \mathrm{ \quad and \quad } \rho = I_X / (I_X + I_{XX}).
\label{eq:g20}
\end{equation}

The red, solid line in Fig. \ref{fig:g2} \textbf{b} depicts a fit to the experimental $g^{(2)}(0)$ values (considering the limited time resolution of the setup) based on equation \ref{eq:g20}. Here we consider $I_X \propto P^{n_X}$ and $I_{XX} \propto P^{n_{XX}}$, as determined by the series of excitation-power-dependent spectra shown in Fig. \ref{fig:stat} \textbf{b} and its inset. A surprisingly good agreement is found between the data and the theoretical treatment assuming an uncorrelated background emission, cf. Fig. \ref{fig:g2} \textbf{b}. Interestingly, in contrast to case (a), the assumption of uncorrelated photon streams related to XX and X does not lead to an underestimation of  $g^{(2)}(0)$ as a function of excitation power density.

The observation that the assumption of an uncorrelated photon stream (case (b)) leads to a better agreement with the experiemental data than the assumption of a correlated photon stream (case (a)), could be understood in terms of the hybrid biexciton cascade depicted in Fig. \ref{fig:schema} \textbf{b}. The spin-flip process between the \emph{dark} and \emph{bright} exciton states should randomize the emission of photons stemming from the overall cascade. However, based on the present measurements, the balance  between the correlated and uncorrelated contributions to the overall background emission, which degrade the $g^{(2)}(0)$ values, remains obscured. Therefore, future measurements should target auto- and cross-correlation analyses for X and XX with variable bandpasses. Clearly, also a better trade-off between $E_{bind}^{XX}$ and the occurring FWHM can be found as suggested by the data shown in Fig. \ref{fig:stat} \textbf{a}.







The loss in single photon purity with increasing excitation power density is accompanied by a narrowing of the antibunching peak as shown in Fig. \ref{fig:g2} \textbf{a}. In Fig. \ref{fig:g2} \textbf{c}, we plot the corresponding FWHM of the antibunching peak as a function of excitation power density. For the considered multiple-level system, this temporal FWHM can be calculated by inverting equation \ref{eq:g2tot}:

\begin{equation}
\tau(g^{(2)}(\tau) = 1/2) = -\tau_d \ln \left( \frac{W(-\tfrac{1}{2} \cdot \mu e^{-\mu})}{\mu}  + 1 \right),
\label{eq:FWHM}
\end{equation}
where $W$ is the Lambert $W$ function. Although a narrowing of the antibunching peak is not inherently detrimental to the SPE performance, it can impact the uncertainty of the measurements as the setup's temporal resolution is approached.

Another salient feature of the $g^{(2)} (\tau)$-measurements is the appearance of a photon bunching phenomenon overlapping with the antibunching signature. Following the analysis of Sallen \emph{et al.} \cite{Sallen2010}, we assume that this bunching is due to spectral diffusion as the emission shifts in and out of the detection window, cf. Fig. \ref{fig:stat} \textbf{b}. With increasing excitation power density the characteristic time of the bunching ($\tau_{SD}$) decreases. This is due to the fact that the rate of spectral diffusion mostly depends  on the occupation probability of charge trapping defects near the QD, which in turn depends on the excitation power density: $1/\tau_{SD} \propto \sqrt{P}$ \cite{Holmes2015, Gao2019}.


Finally, $g^{(2)} (\tau)$ measurements were performed on several QDs at 300 K with an excitation power density of 20 kW cm$^{-2}$. QDs with an emission energy between 4.4 and 4.6 eV and large $E_{bind}^{XX}$ values were selected in order to maintain the validity of the previous discussion. All measured QDs displayed antibunching (see supplementary material). The average $g^{(2)} (0)$ value amounted to 0.23 $\pm$ 0.05. The high yield of GaN QDs displaying SPE at 300 K together with the high count rates and the use of Si substrates pave a promising pathway towards future integration of these QDs into  integrated photonic platforms. For example, integration of a QD into an optical cavity has been shown to increase the indistinguishability of the photons as it decreases phonon decoherence processes which are particularly pronounced at elevated temperatures \cite{Grange2017}. Further on, quantum frequency down-conversion schemes such as the one presented by Zaske \emph{et al.} \cite{Zaske2012}, enable the conversion of QD based single photon sources to telecom wavelengths.


\section*{Summary}

In conclusion we have demonstrated that self-assembled GaN/AlN QDs on Si substrates can provide SPE at 300 K. Their spectral signature and their behavior with temperature can be explained by a hybrid XX model and strong exciton-phonon interaction. These QDs remain bright even at room temperature, with typical count rates at the sample surface on the order of 10$^6$ - 10$^7$ s$^{-1}$. Photon statistics displayed an enhanced single photon purity for low excitation powers due to the suppressed contribution of the XX. The overlapping bunching in the $g^{(2)} (\tau)$ is explained in terms of spectral diffusion. The probed QDs exhibited SPE at 300 K with an average value $g^{(2)} (0) = 0.23$ and a minimum value of $g^{(2)} (0) = 0.17 \pm 0.08$.


\subsection*{Methods}

The sample was grown by NH$_3$-molecular beam epitaxy (MBE). It consists of 100 nm of high temperature AlN on Si(111), followed by a plane of GaN QDs, a 20 nm AlN top barrier and finally a second plane of GaN QDs. Further details on the growth of the AlN layer and the QDs can be found in Refs. \cite{Tamariz2017, Tamariz2019}. The top QD plane  was used for atomic force microscopy (AFM) and subsequently evaporated in the MBE reactor under vacuum at a temperature of 800 $^\circ$C. Mesa structures and position markers were then fabricated by e-beam lithography. The developed regions of the resist were etched down to a depth of about 60 nm, so that no QDs remain outside of the mesa and marker structures.  The mesa sizes range from 2 $\mu$m to 50 nm in diameter.  From AFM measurements, the QD density was estimated to be $\approx$ 10$^{10}$ cm$^{-2}$ (see supplementary material). Furthermore, AFM revealed a broad QD size distribution,  which is advantageous for the study of the optical features of QDs emitting over a wide range of energies.

The optical properties were investigated by $\mu$-PL spectroscopy. The QDs were excited either with a 266 nm or a 244 nm continuous wave laser. The sample has a very thin wetting layer, estimated at a thickness of 1.5 ML, that is not excited with the employed lasers. The sample was loaded in a closed-cycle He cryostat (Cryostation C2 from Montana Instruments, Inc.) and the luminescence was collected by an 80$\times$ infinity corrected microscope objective suited for the ultraviolet spectral range (NA = 0.55). The laser spot size was estimated to have a diameter of  $\approx$ 1 $\mu$m. The luminescence is then routed either to a high resolution spectrometer or a Hanbury Brown and Twiss interferometer for $g^{(2)} (\tau)$ measurements.

\subsection*{Supplementary material}
See supplementary material for additional structural  and  spectroscopic information, as well as the treatment of correlated emission in the $g^{(2)}(\tau)$ measurements.

\subsection*{Acknowledgements}
This work was supported by the Swiss National Science Foundation through Grant 200021E\_15468 and 200020\_162657 and by the Marie Sklodowska-Curie action ``PhotoHeatEffect" (Grant No. 749565) within the European Union's Horizon 2020 research and innovation program.

\subsection*{Author contribution}

S.T., G.C. and J.S. performed the optical measurements. S.T. grew the sample and  K.S. performed the nano-mesa fabrication. S.T., G.C., J.S. and R.B. analyzed the data. S.T. and G.C. wrote the manuscript with contributions from all authors. N.G. initiated and supervised the entire project.


	

\bibliography{RT_GaN_ArXiv}

\end{document}


\title{Supplementary Material: Towards bright and pure single photon emitters at 300 K based on GaN quantum dots on silicon}

\author{Sebastian Tamariz}
\email{sebastian.tamariz@epfl.ch}
\affiliation{Institute of Physics, \'Ecole Polytechnique F\'ed\'erale de Lausanne, EPFL, CH-1015 Lausanne, Switzerland}
\author{Gordon Callsen}
\affiliation{Institute of Physics, \'Ecole Polytechnique F\'ed\'erale de Lausanne, EPFL, CH-1015 Lausanne, Switzerland}
\author{Johann Stachurski}
\affiliation{Institute of Physics, \'Ecole Polytechnique F\'ed\'erale de Lausanne, EPFL, CH-1015 Lausanne, Switzerland}
\author{Kanako Shojiki}
\affiliation{Institute of Physics, \'Ecole Polytechnique F\'ed\'erale de Lausanne, EPFL, CH-1015 Lausanne, Switzerland}
\affiliation{Current address: Graduate School of Engineering, Mie University, Tsu 514-8507, Japan}
\author{Rapha\"el Butt\'e}
\affiliation{Institute of Physics, \'Ecole Polytechnique F\'ed\'erale de Lausanne, EPFL, CH-1015 Lausanne, Switzerland}
\author{Nicolas Grandjean}
\affiliation{Institute of Physics, \'Ecole Polytechnique F\'ed\'erale de Lausanne, EPFL, CH-1015 Lausanne, Switzerland}

\maketitle

\subsection{Sample description}

\begin{figure*}[h]
	\centering
	\includegraphics[width=14cm]{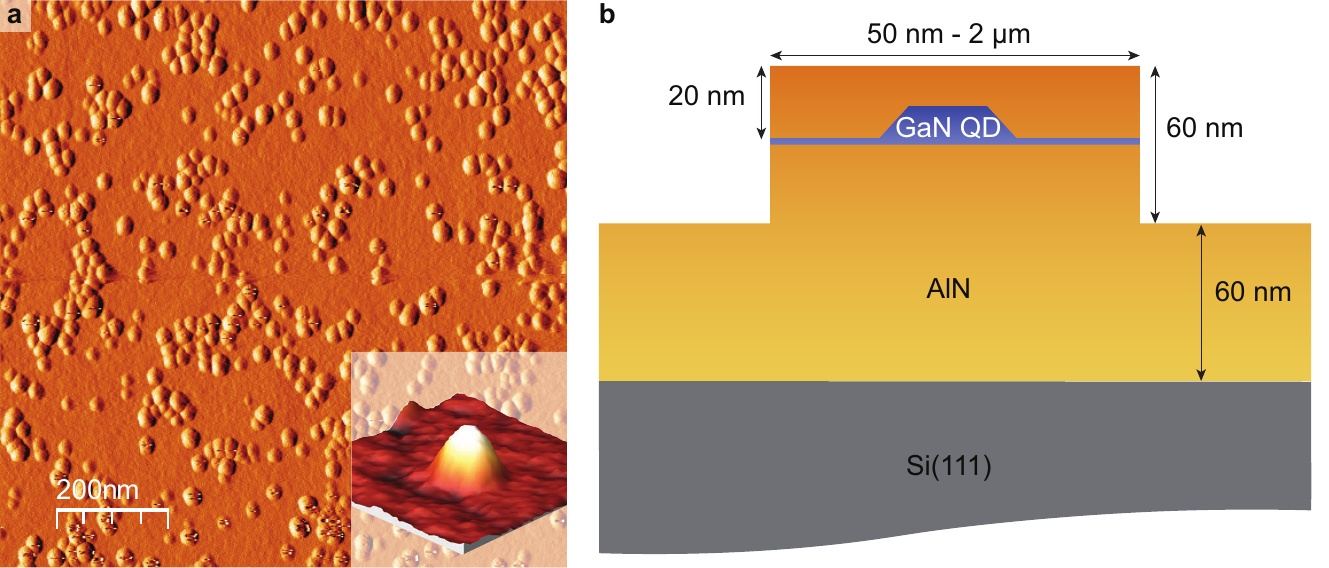}
	\caption{\textbf{a} 1$\times$1 $\mu$m$^2$ AFM image in amplitude mode of the top plane of GaN QDs displaying a broad distribution of QD sizes. Inset: three dimensional render of a QD taken from the same AFM scan. \textbf{b} Schematic cross section of the sample structure after evaporation of the GaN QD top plane and mesa fabrication. }
	\label{fig:s1AFM}
\end{figure*}

The sample structure consists of 100 nm of AlN on Si(111), one plane of GaN quantum dots (QDs) with a 20 nm AlN top barrier and a top plane of GaN QDs for atomic force microscopy (AFM) imaging. The AFM image of the sample (Fig. \ref{fig:s1AFM} \textbf{a}) exhibits clear three dimensional islanding and a broad distribution of QD sizes. The density was estimated to $\approx$ 10$^{10}$ cm$^{-2}$. After AFM, the top plane of GaN QDs was evaporated in the molecular beam epitaxy (MBE) chamber; GaN is unstable in vacuum for substrate temperatures above 750 $^\circ$C, whereas AlN does not evaporate up to temperatures as high as 1200 $^\circ$C \cite{Grandjean1999, Tamariz2017}. Hence, surface GaN can be evaporated selectively from AlN. Mesa fabrication was performed by e-beam lithography after evaporation of the top plane of GaN QDs. The process employed a SiO$_2$ hard mask, ZEP-520A positive resist, and chlorine-based inductively coupled plasma etching. In Fig. \ref{fig:s1AFM} \textbf{b}, a schematic cross section of the sample is displayed after the evaporation of the top GaN QD plane and subsequent mesa fabrication.  

\newpage

\subsection{Optical characterization methods}

\begin{figure}[h]
	\centering
	\includegraphics[width=10cm]{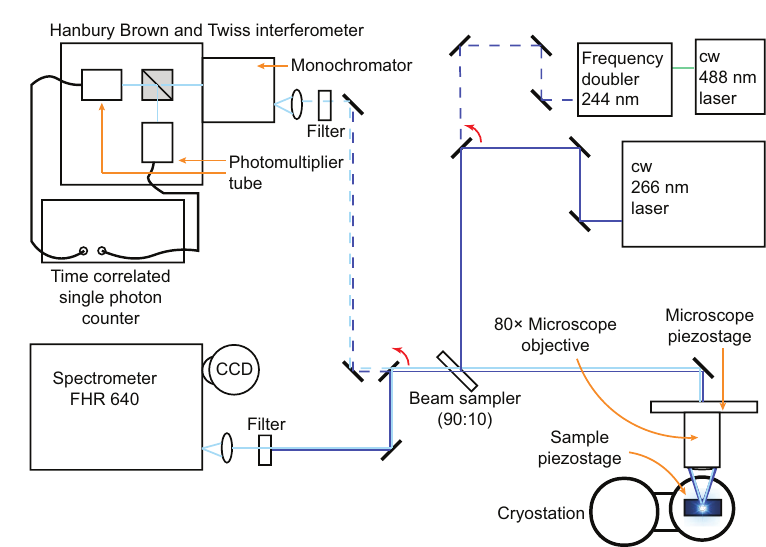}
	\caption{Schematic illustration of the $\mu$-PL setup employed for this work. The red curved arrows signal flip mirrors and the dashed lines alternative optical paths.}
	\label{fig:uPLsetup}
\end{figure}

As continuous wave excitation sources we either employed a frequency-doubled 488\,nm laser (Coherent Genesis CX SLM laser pumping a Spectra-Physics WaveTrain frequency doubler), resulting in an excitation wavelength of 244 nm (5.08\,eV), or a 266\,nm laser from Crylas (4.66\,eV). The laser was guided toward the sample via a 90:10 (transmission/reflection) beamsplitter and a microcope objective (80$\times$ Mitutoyo Plan UV Infinity Corrected Objective) as illustrated in Fig.\,\ref{fig:uPLsetup}. 

The sample was placed in a closed-cycle helium cryostat (Cryostation C2 from Montana Instruments, Inc.) allowing temperature variations from 5 to 300\,K. Fine mappings of the sample surface were achieved by displacing the microscope objective via a piezostage (Physik Instrumente P-612.2 XY). The luminescence of the individual QDs was routed backwards via the same beam path toward the beamsplitter. Mirrors mounted on flipmounts allowed selecting between two different detection pathways: Standard photoluminescence acquisition was done by using a single monochromator (Horiba FHR 640, 1800\,l/mm holographic  grating blazed at 400\,nm) in combination with a liquid nitrogen cooled CCD (Horiba Symphony II, UV-enhanced). The resolution of this detection pathway was chosen to be better than 900\,$\mu$eV for the entire QD emission interval. The second detection pathway, as illustrated by the dashed lines in Fig.\,\ref{fig:uPLsetup}, comprises a single monochromator (SPEX 270M, holographic 2400\,l/mm grating blazed at 250\,nm) attached to a Hanbury Brown and Twiss (HBT) interferometer for autocorrelation ($g^{(2)}(\tau)$) measurements. The bandpass of this second detection pathway was chosen to be $\approx$ 8\,meV. The autocorrelator featured a 50:50 beamsplitter and two photomultiplier tubes (PicoQuant PMA 175) connected to a time-correlated single photon counter (PicoQuant, PicoHarp 300). The measured bi-photon time resolution of the autocorrelation setup yielded 220\,ps.

The overall efficiency of both detection pathways was carefully examined. Hereunto, the 4.66\,eV laser was coupled through the entire optical setup depicted in Fig.\,\ref{fig:uPLsetup}. A powermeter was used to determine the optical losses of each individual element in the beam path. As a result, the setup efficiency was measured at 4.66\,eV and subsequently extrapolated to $\approx$ 4.5\,eV via the documented energy dependence of, e.g., the transmission, reflection (optical elements), and detection efficiency (photomultiplier tubes) of the relevant elements. The overall efficiency of the second detection pathway was estimated to 0.11 $\pm$ 0.04 \% at an energy of $\approx$ 4.5\,eV.


\subsection{Further optical measurements}

\begin{figure*}
	\centering
	\includegraphics[width=0.98\textwidth]{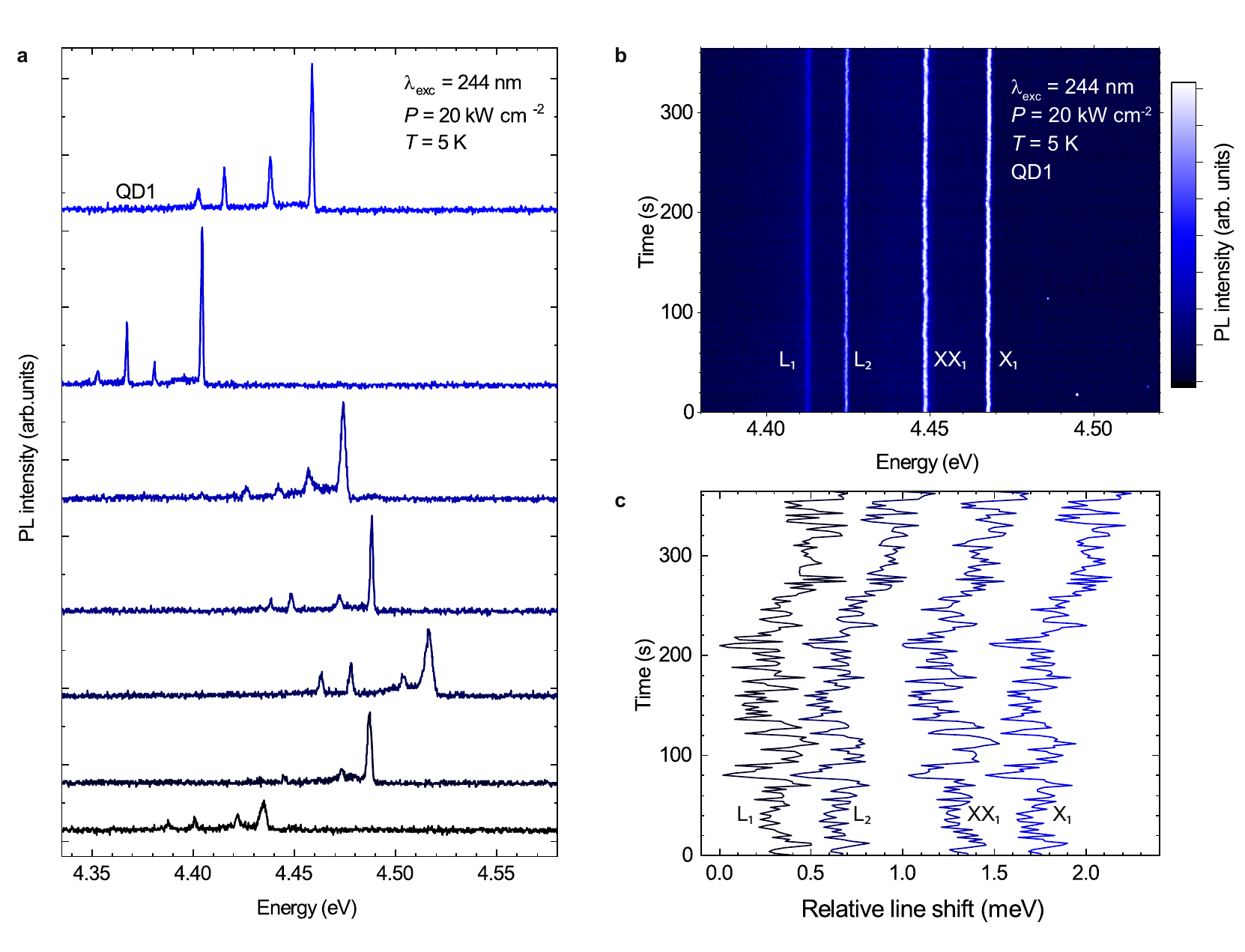}
	\caption{\textbf{a} Low temperature $\mu$-PL spectra of several GaN QDs emitting at an energy around 4.4 eV. The spectra are dominated by four peaks and the highest energy transition is the most intense. \textbf{b} Time evolution of the $\mu$-PL emission of QD1, fluctuations in the emission energy due to spectral diffusion. \textbf{c} Relative energy shift of the lines shown in \textbf{b} (horizontally shifted for clarity), all the lines shift at the same moment in time and with similar amplitude.} 
	\label{fig:s3PL}
\end{figure*}

GaN QDs emitting at energies around 4.4 eV often display a similar optical signature to QD1 (see main text), where four lines dominate the emission spectrum and the highest energy line is the brightest. Several examples of the emission of single GaN QDs emitting at similar energies as QD1 are displayed in Fig. \ref{fig:s3PL} \textbf{a}. The similarity among these spectra led us to conclude that QD1 is representative of QDs emitting at energies around 4.4 eV. Further on, similar spectra, with the presence of L$_1$ and L$_2$ were also observed in GaN QDs by Kindel \emph{et al.} \cite{Kindel2010} and in InGaN QDs by Amloy \emph{et al.} \cite{Amloy2011}. The recursive appearance of such optical footprint suggests that each set of lines originates from a single given QD. Further on, spectral diffusion can be used to group spectral lines originating from the same QD \cite{Bardoux2006}. Defects near a QD can stochastically trap and release electrical charges resulting in a fluctuating emission energy. Then,  variations of the local environment are spatially specific to each QD. Hence, any excitonic complex within the QD will display similar changes in emission energy \cite{Honig2013}. In Fig. \ref{fig:s3PL} \textbf{b} the temporal evolution of the micro-photoluminescence ($\mu$-PL) emission of QD1 is plotted together with the relative line shift of each emission line in Fig \ref{fig:s3PL} \textbf{c} (sampling rate of 2 s). The synchronous change of the four dominant emission lines of QD1 further indicates that all the lines originate from the same QD.

\begin{figure}[h]
	\centering
	\includegraphics[width=10cm]{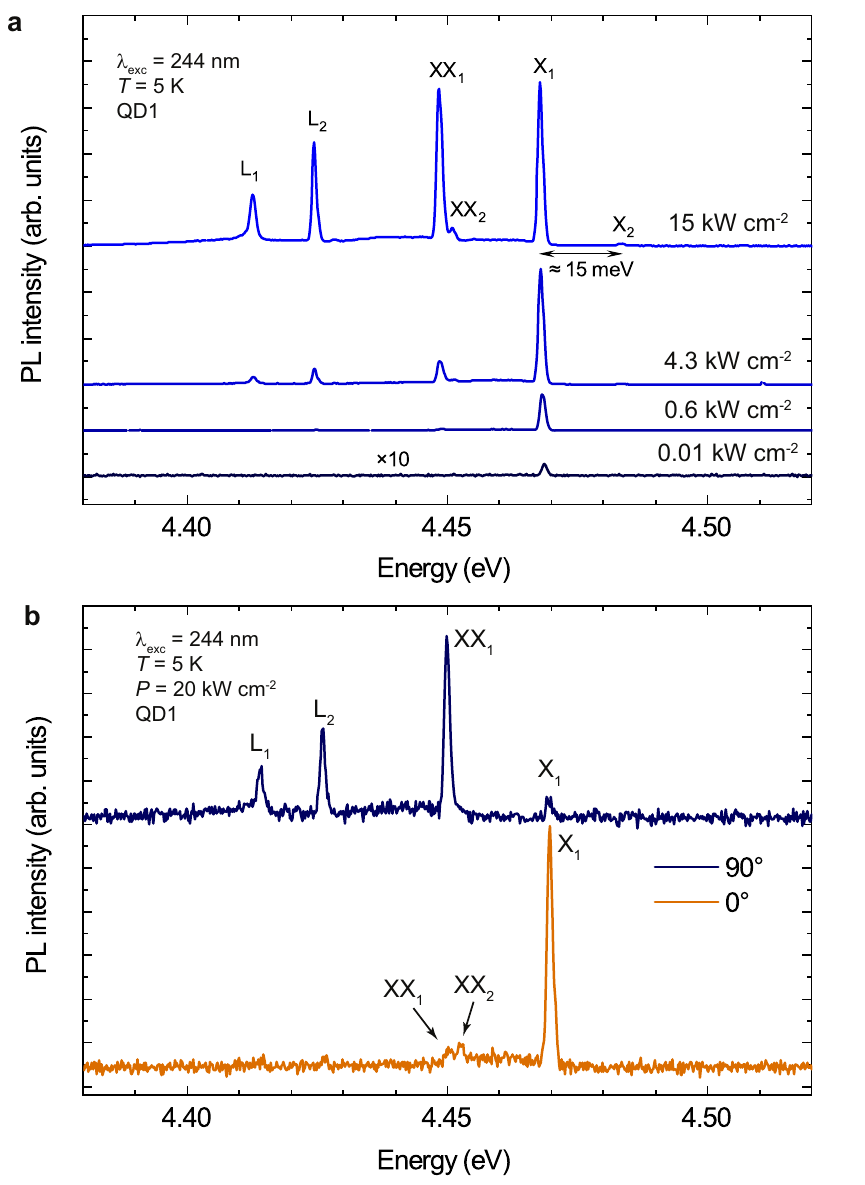}
	\caption{\textbf{a} Low temperature power-dependent $\mu$-PL spectra of QD1 (vertically shifted for clarity). At the lowest excitation power density, only the exciton is visible. \textbf{b} Low temperature $\mu$-PL spectra of QD1 for orthogonal polarizations (vertically shifted for clarity). All transitions are linearly polarized, with XX$_2$ and X$_1$ being cross-polarized to all other spectral lines.}
	\label{fig:s4Pwrpol}
\end{figure}

The identification of the exciton line in QD1 is supported by power-dependent measurements shown in Fig. \ref{fig:s4Pwrpol} \textbf{a}. The X$_1$ line is singled out by lowering the excitation power density. Similarly to QD2 (see main text), the intensity of the transitions as a function of excitation power density can be modeled by a Poisson distribution, yielding $n_{\mathrm{L_1}}$ = 1.50 $\pm$ 0.04, $n_{\mathrm{L_2}}$ = 1.47 $\pm$ 0.04, $n_{\mathrm{XX_1}}$ = 1.43 $\pm$ 0.02 and $n_{\mathrm{X_1}}$ = 0.90 $\pm$ 0.02 along the lines explained in the main text. The identification of the transitions is further supported by polarization-resolved spectra displayed in Fig. \ref{fig:s4Pwrpol} \textbf{b}: XX$_1$ and X$_2$ (not visible) are cross-polarized to XX$_2$ and X$_1$ \cite{Honig2014, Arita2017}. A similar polarization pattern was found by H\"onig \emph{et al.} \cite{Honig2014} and Arita \emph{et al.} \cite{Arita2017}  for GaN/AlN QDs and GaN/Al$_{0.2}$Ga$_{0.8}$N interface fluctuation QDs, respectively. However, the energetic splitting between the biexcitonic (XX$_1$ and XX$_2$) and excitonic states (X$_1$ and X$_2$) strongly deviates among these reports due to the occurrence of the hybrid biexciton in GaN/AlN QDs and a conventional biexciton in the GaN/Al$_{0.2}$Ga$_{0.8}$N interface fluctuation QDs. Furthermore, the emission lines L$_1$ and L$_2$ are cross-polarized to X$_1$ and XX$_2$.

\newpage

\begin{figure}[h]
	\centering
	\includegraphics[width=10cm]{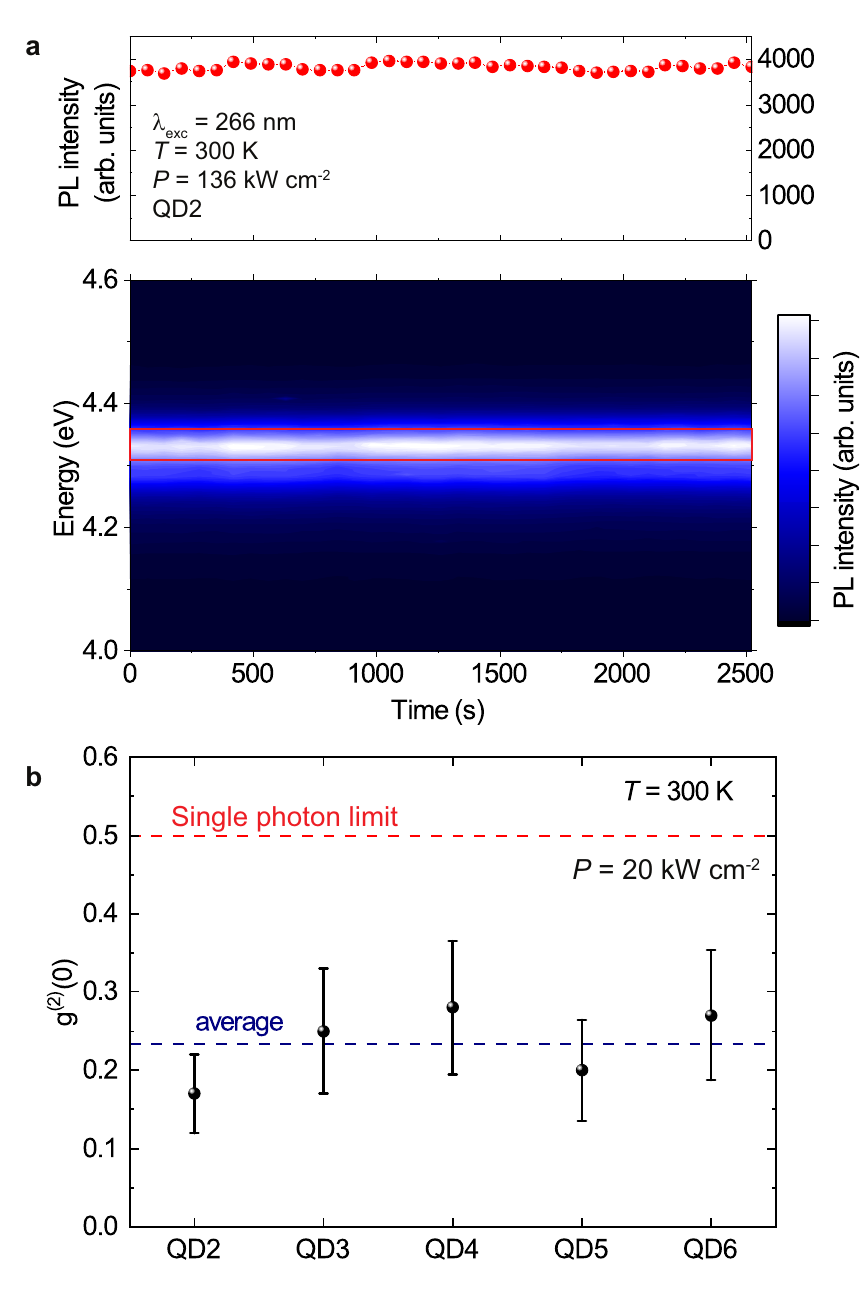}
	\caption{\textbf{a} Long term PL stability of QD2 under constant excitation at room temperature.  \textbf{b} $g^{(2)}(0)$ value of the measured QDs at room temperature.}
	\label{fig:s5time_g2}
\end{figure}

Figure \ref{fig:s5time_g2} \textbf{a} highlights the longterm time stability (2500\,s) of the QD emission based on QD2, which we commonly observe at 300\,K. This excellent emission stability is accompanied by a constant QD emission energy, cf. Fig. \ref{fig:s5time_g2} \textbf{a}. Furthermore, no blinking was observed in any of the QDs. As a result of this overall emission stability, we were able to record $g^{(2)}(\tau)$-traces for five QDs as shown in Fig. \ref{fig:s5time_g2} \textbf{b}. The excitation power density was set to 20 kW cm$^{-2}$ and the sample temperature to 300 K (except for QD2 where we report the $g^{(2)}(0)$ obtained with $P$ = 6.5 kW cm$^{-2}$). All probed QDs showed a pronounced antibunching in the $g^{(2)}(\tau)$ function.  Here, we determine an average $g^{(2)}(0)$ of 0.23 $\pm$ 0.05 for five QDs.

\subsection{Analysis of the single photon purity as a function of excitation power density}

\begin{figure}[h]
	\centering
	\includegraphics[width=10cm]{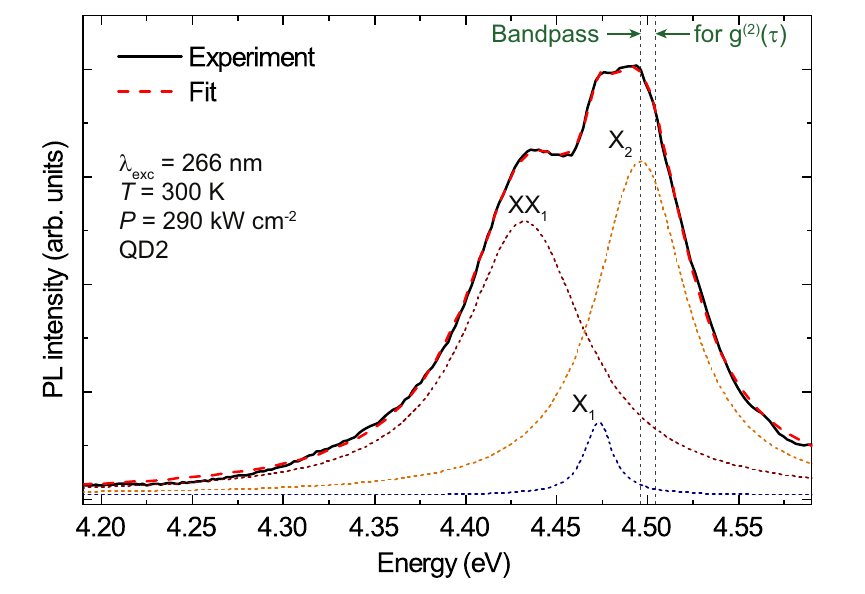}
	\caption{Room temperature $\mu$-PL spectrum of QD2  the corresponding fit employing Lorentzian line shapes. A clear overlap of the X$_2$ and XX$_1$ transitions can be observed in the bandpass used for the $g^{(2)}(\tau)$ measurements.}
	\label{fig:S6QD2}
\end{figure}

Exciton and biexciton emission from the same quantum dot are expected to be temporally correlated. As observed when performing cross-correlation measurements of these transitions \cite{Moreau2001, Kiraz2002, Sallen2009}: the detection of the biexciton leads to an increased emission probability for the exciton resulting in  antibunching followed by  bunching. In our experiments, there is an overlap of the biexciton and the exciton emission within the selected bandpass for the $g^{(2)}(\tau)$ measurements, cf. Fig. \ref{fig:S6QD2}. The impact of the overlap in the $g^{(2)}(\tau)$ function, will be  described by first considering the absence of any spectral filtering and then adapted to the intensities of each transition within the bandpass.


Without any spectral filtering, one has to consider all correlation channels, i.e., all possible auto- and cross-correlation functions:
\begin{equation} 
g^{(2)}_{tot}(\tau) \propto \sum_{i,j} \gamma_i \gamma_j p_i p_j \cdot g^{(2)}_{i\rightarrow j}(\tau) ,
\label{eq:allcorr}
\end{equation}
which involves the decay rates $\gamma_{i,j}$ and occupation probabilities $p_{i,j}$ of the $i^{\mathrm{th}}$ and $j^{\mathrm{th}}$ excitonic states. The average emission rate of photons ($\overline{n}_i$) is then given by $\overline{n}_i = \gamma_{i} \cdot p_i$ \cite{Moreau2001}. In our case, it is convenient to write equation \ref{eq:allcorr} as a function of PL intensity: $I_i = \overline{n}_i\cdot T_M \cdot \kappa$,  where $T_M$ is the integration time  and $\kappa$ is a scaling factor,  which takes into account the width of the $g^{(2)}$ bandpass as well as the detection losses: 

\begin{equation}
g^{(2)}_{tot}(\tau) = C \cdot \sum_{i,j} I_i I_j \cdot g^{(2)}_{i\rightarrow j}(\tau), 
\end{equation}
where $C$ is a normalization constant. By definition $\forall \ i,j \ \ \lim\limits_{\tau \rightarrow \pm \infty} g^{(2)}_{i\rightarrow j}(\tau) = 1$ and $\lim\limits_{\tau \rightarrow \pm \infty} g^{(2)}_{tot}(\tau) = 1$, from which it follows that:

\begin{equation}
C = 1/ \sum_{i,j} I_i I_j.
\end{equation}

For the particular case of QD2, only the biexciton and the exciton transitions are observed in the PL spectra. Hence, $C$ can be readily calculated:

\begin{equation}
C  = 1/ \sum_{i,j=X,XX} I_i I_j =  1/ ( I_X^2 + 2 I_X I_{XX} + I_{XX}^2).
\end{equation}

The impact of the presence of the biexciton and the exciton in our detection bandpass on the single photon purity ($g^{(2)} (0)$) can be determined by evaluating the function at zero time delay. The value of $g^{(2)} (0)$ vanishes for the auto-correlation of X and XX and the cross-correlation X$\rightarrow$XX. The only non-vanishing term is the cross correlation XX$\rightarrow$X \cite{Kindel_thesis}:

\begin{equation}
g^{(2)}_{XX\rightarrow X}(\tau) = \exp(\mu \cdot e^{- \mid \tau \mid}/ \tau_d) \cdot ( (1- e^{- \mid \tau \mid / \tau_d})^2 +1/\mu \cdot e^{- \mid \tau \mid / \tau_d}),
\label{eq:cross}
\end{equation}
where $\mu = \Pi \cdot \tau_d$, with $\tau_d$ the decay time and $\Pi$ the pump rate. After evaluating equation \ref{eq:cross} at zero time delay, the single photon purity can be determined: 

\begin{equation}
g^{(2)}_{tot}(0) = C \cdot I_X I_{XX} \cdot e^\mu/\mu = \frac{1}{2 + I_X/I_{XX} + \cdot I_{XX}/I_X} \cdot e^\mu / \mu.
\label{eq:long}
\end{equation}

The intensity of each spectral line can be directly extracted from $\mu$-PL measurements, as exemplified in Fig. \ref{fig:S6QD2}, and $\mu$ is ascertained by fitting the coincidence histograms displayed in the main text (Fig. 4 \textbf{a}).

\bibliography{SI-RT_GaN_ArXiv}